\documentclass[letterpaper, 10 pt, conference]{ieeeconf}  

\IEEEoverridecommandlockouts                              

\def\pr{\mathrm{Pr}}

\title{\LARGE \bf Probabilistically  Certified Region of Attraction of a Tumor Growth Model with Combined Chemo- and Immunotherapy
}

\author{
Kaouther Moussa$^{*}$, Mirko Fiacchini$^{*}$ and Mazen Alamir$^{*}$
        \thanks{$^{*}$ Univ. Grenoble Alpes, CNRS, Grenoble INP, GIPSA-lab, 38000 Grenoble, France
        {\tt\small \{kaouther.moussa,mirko.fiacchini,}
        {\tt\small mazen.alamir\}@grenoble-inp.fr}}%
}

\usepackage[T1]{fontenc}

\usepackage[utf8]{inputenc} 
\usepackage{amsmath}
\usepackage{tikzscale}
\usepackage{pgfplots}
\usepackage{tikz}
\usepackage[np]{numprint}
\usepackage{times}
\usepackage{soulpos}
\usepackage{xspace}
\usepackage[lofdepth=1,lotdepth]{subfig}
\usepackage{caption}
\usepackage{lmodern}
\usepackage{graphicx}
\usepackage[colorinlistoftodos]{todonotes}
\usepackage{algorithm}
\usepackage{algorithmic}
\usepackage{amssymb}
\usepackage{booktabs}
\usepackage{color,soul}
\usetikzlibrary{calc,trees,positioning,arrows,chains,shapes.geometric,%
    decorations.pathreplacing,decorations.pathmorphing,shapes,%
    matrix,shapes.symbols}

\usepackage{lipsum}

\def\R{\mathbb{R}}

\definecolor{red}{rgb}{0.76, 0.13, 0.28} \definecolor{bt}{rgb}{0.03, 0.91, 0.87}
\definecolor{bo}{rgb}{0.8, 0.33, 0.0}
\definecolor{cardinal}{rgb}{0.77, 0.12, 0.23}
\definecolor{coolblack}{rgb}{0.0, 0.18, 0.39}
\definecolor{darkblue}{rgb}{0.0, 0.0, 0.55}
\definecolor{mycolor1}{rgb}{0.00000,0.44700,0.74100}
\definecolor{forestgreen}{rgb}{0.0, 0.27, 0.13}
\definecolor{persiangreen}{rgb}{0.0, 0.65, 0.58}
\definecolor{cblue}{rgb}{0.16, 0.32, 0.75}
\definecolor{burntumber}{rgb}{0.54, 0.2, 0.14}
\definecolor{cadetblue}{rgb}{0.37, 0.62, 0.63}
\definecolor{burntsienna}{rgb}{0.91, 0.45, 0.32}
\definecolor{cadet}{rgb}{0.33, 0.41, 0.47}
\definecolor{maroon}{rgb}{0.69, 0.19, 0.38}
\definecolor{byzantium}{rgb}{0.44, 0.16, 0.39}
\definecolor{polypomonagreen}{rgb}{0.12, 0.3, 0.17}
\definecolor{darkgoldenrod}{rgb}{0.72, 0.53, 0.04}
\definecolor{debianred}{rgb}{0.84, 0.04, 0.33}
\definecolor{mint}{rgb}{0.24, 0.71, 0.54}
\definecolor{oldlavender}{rgb}{0.47, 0.41, 0.47}
\definecolor{paleviolet-red}{rgb}{0.86, 0.44, 0.58}
\definecolor{raspberryrose}{rgb}{0.7, 0.27, 0.42}
\definecolor{darkmagenta}{rgb}{0.55, 0.0, 0.55}
\definecolor{benign}{rgb}{0.47059,0.67059,0.18824}%
\definecolor{malignant}{rgb}{0.92941,0.69020,0.12941}%
\definecolor{ikb}{rgb}{0.0, 0.18, 0.65}
\definecolor{bazaar}{rgb}{0.6, 0.47, 0.48}
  \definecolor{bole}{rgb}{0.47, 0.27, 0.23}
\definecolor{darkcerulean}{rgb}{0.91, 0.33, 0.5}
\definecolor{bittersweet}{rgb}{1.0, 0.44, 0.37}
\definecolor{copper}{rgb}{0.72, 0.45, 0.2}
\definecolor{darkcoral}{rgb}{0.8, 0.36, 0.27}
	\definecolor{midnightgr}{rgb}{0.0, 0.29, 0.33}

\newtheorem{theorem}{\bf{Theorem}}

\renewcommand{\algorithmicrequire}{\textbf{Input:}}
\renewcommand{\algorithmicensure}{\textbf{Output:}}

\begin{document}

\maketitle

\begin{abstract}
This paper deals with the estimation of regions of attraction (RoAs) under parametric uncertainties for  a cancer growth model with combined therapies. We propose a framework of probabilistic certification, based on the randomized methods, in order to derive probabilistically certified RoAs of a cancer growth model.  The model that we consider in this paper describes the interaction between tumor and immune system in presence of a combined chemo- and immunotherapy. Furthermore, we model the concentration of the chemotherapy agent in the body via a pharmacokinetic equation.
\end{abstract}

\section{Introduction}
The last decades witnessed a considerable progress in experimental and clinical immunology \cite{Eftimie2016} as well as in modeling the immune system dynamics. 

The progress in cancer dynamics modeling motivated researchers to apply control approaches in order to schedule cancer treatments using optimal control strategies. We can find in the literature many works regarding the application of optimal control approaches on cancer treatment problems. For instance \cite{Dono2009a}, where optimal protocols for anti-angiogenic therapy were investigated, or \cite{DePillis2007} where  linear controls were designed for a tumor-immune interactions model with chemotherapy delivery. However, only few works addressed the problem of handling parametric uncertainties. One can cite for example, \cite{Alamir2014} where a robust feedback scheme is proposed to schedule antiangiogenic treatment combined with chemotherapy,  \cite{Kovacs2014} where an $H_{\infty}$-based robust control was applied to the same model and \cite{Alamir2015} where a general framework for probabilistic certification of cancer therapies was proposed.

The estimation of the region of attraction for cancer models is an interesting 
problem since it provides a set of possible initial conditions (tumor 
volume and immune density for example) that can be driven to a desired target 
set (benign region). This problem becomes complex when dealing with  nonlinear 
systems and even more challenging for uncertain systems. There are some works 
which dealt with the problem of estimating the RoA for cancer models but only 
few of them considered model uncertainties.  In particular, in \cite{Riah2019}, 
an iterative method to estimate the robust RoA was presented. However, 
robust RoA estimation is based on the worst-case scenario analysis leading to 
very pessimistic design. This is because the worst case is considered no matter how 
small its probability of occurrence is. 

In this paper, we propose a framework to 
probabilistically certify the existence of a control structure that drives the 
states corresponding to tumor cells and immune density from an initial state set 
to a certified target set. This probabilistic certification framework is based 
on the randomized methods proposed in \cite{Alamo2009} and \cite{Alamo2015}, 
which, unlike the robust classical design, avoids focusing on few unlikely very 
bad scenarios allowing to overcome the conservatism of the robust RoA design. The methodology that we propose in this paper consists mainly of two steps. Firstly, we derive an ordered sequence of sets and a control strategy over each of them such that the states can be driven from a set to the previous with a certain probabilistic guarantee. The appropriate choice of the first set allows to insure that the union of the sets is a probabilistically certified approximation of the RoA. The second step consists of providing a global certification on the probability of convergence to the initial certified target set.

This paper is organized as follows: In Section~II, the dynamical cancer model and the problem of RoA 
probabilistic certification are introduced. In Section~III,  a framework for RoA probabilistic  certification is proposed, based on the randomized methods presented in \cite{Alamo2009} and 
\cite{Alamo2015}. In Section~IV, the proposed RoA probabilistic certification framework is applied to the considered cancer model. Finally, Section~V summarizes the contribution and gives some hints for further investigation.

\section{Problem statement}

The following nonlinear dynamical system describes the interaction between 
tumor and  immune system in presence of chemotherapy and immunotherapy drugs:
\begin{equation}
\begin{aligned}
\dot{x}_1&=\mu_C x_1 -\dfrac{\mu_C}{x_{\infty}}x_1^2-\gamma x_1 x_2 - \sigma x_1 x_3,  \\
\dot{x}_2&= \mu_I x_1 x_2 - \beta x_1^2 x_2 - \chi x_2 + \lambda x_2 u_2 -\varrho x_3 x_2 + \alpha,\\
\dot{x}_3&=-a x_3 + b u_1,\\
x(0)&= \left( x_1(0),x_2(0),x_3(0) \right)=x_0,
\end{aligned}
\label{Model2}
\end{equation}
where $x_1$, $x_2$ and $x_3$ denote, respectively, the number of tumor cells, 
the density of effector immune cells (ECs) and the concentration of chemotherapy 
in the body, $u_1$ and $u_2$ are, respectively, the dosages of a cytotoxic agent 
and an immuno-stimulator.  This model gives the advantage of a low dimensional system that nevertheless includes  the main aspects of cancer-immune interactions.

In many models it is assumed that the drug concentration is equal to its dosage 
which is an oversimplification. Therefore, we revisited the model proposed in 
\cite{DOnofrio2012} by adding a pharmacokinetic (PK) equation that allows to 
model the concentration of chemotherapy in the body. This equation is a 
classical PK model with an exponential  growth/decay of the drug concentration.

Fig.~\ref{Model_Scheme} presents a scheme describing the different interactions 
between the tumor and the immune system. Table~\ref{Table2} summarizes the 
definitions of the model parameters and their numerical values. We slightly 
tuned the values of some parameters since with the previous set of parameters 
values (used in \cite{ACC2020} and \cite{FOSBE2019}), the domain of attraction 
of the benign equilibrium for the uncontrolled system~(\ref{Model2}) (for $u_1=0$ and $u_2=0$) was unrealistically big, this allows us to solve a more challenging and seemingly realistic problem. Moreover, we properly 
chose the parameters $a$ and $b$ of the PK dynamics, such that the drug 
concentration reaches its maximum in $4.8h$ and  starts decreasing towards a 
negligible value after a period of $15$ days. Nevertheless, it is worth emphasizing that in this paper, we focus on the 
assessment of a methodology that remains applicable for different nominal and PK 
parameters values. 
\definecolor{red}{rgb}{0.76, 0.13, 0.28} \definecolor{bt}{rgb}{0.03, 0.91, 0.87}
\definecolor{bo}{rgb}{0.8, 0.33, 0.0}
\definecolor{cardinal}{rgb}{0.77, 0.12, 0.23}
\definecolor{coolblack}{rgb}{0.0, 0.18, 0.39}
\definecolor{darkblue}{rgb}{0.0, 0.0, 0.55}
\definecolor{mycolor1}{rgb}{0.00000,0.44700,0.74100}
\definecolor{forestgreen}{rgb}{0.0, 0.27, 0.13}
\definecolor{persiangreen}{rgb}{0.0, 0.65, 0.58}
\definecolor{cblue}{rgb}{0.16, 0.32, 0.75}\definecolor{burntumber}{rgb}{0.54, 0.2, 0.14}

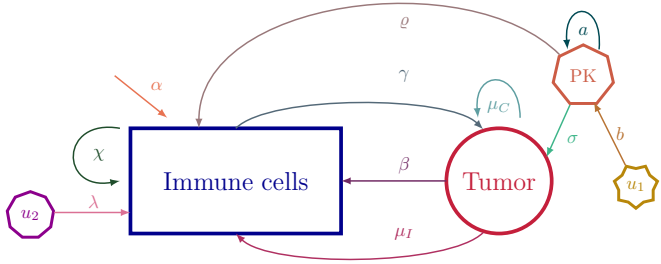
\begin{figure}
\centering
\scalebox{0.7}{
\begin{tikzpicture}
\tikzstyle{suite}=[->,>=stealth’,thick,rounded corners=4pt]
\tikzstyle{operation}=[->,>=latex,thick]
\draw[line width=2.0pt,cardinal] (0,0) circle   (1cm) ;
\node[cardinal,ultra thick] (tumor) at (0,0) {\Large $\text{Tumor}$ };
\draw[line width=2.0pt,darkblue]  (-3,1) rectangle   (-7,-1) ;
\node[darkblue,ultra thick] (immune) at (-5,0) {\Large $\text{Immune cells}$ };
\draw[operation,cadetblue] (0.4,1.2) to[out=90,in=95, looseness=3] (-0.4,1.2) node[cadetblue] at (0,1.4)  {$\mu_C$};

\draw[operation,burntsienna] (-7.3,2) -- (-6.3,1.2) node[burntsienna] at (-6.5,1.8)  {$\alpha$}; 
\draw[operation,cadet] (-5,1) to[in=135,out=45, looseness=0.5] (-0.3,0.98) node[cadet] at (-1.8,2)  {$\gamma$}; 
\draw[operation,maroon] (-0.3,-0.98) to[in=-45,out=-135, looseness=0.5](-5,-1)  node[maroon] at (-1.8,-1)  {$\mu_I$}; 
\draw[operation,byzantium] (-1,0) -- (-3,0) node[byzantium] at (-1.8,0.3)  {$\beta$}; 

\draw[operation,polypomonagreen] (-7.2,1) to[out=170,in=190, looseness=3] (-7.2,0) node[polypomonagreen] at (-7.6,0.5)  {$\chi$};

\draw[operation,bazaar] (1.15,2.4) to[in=90,out=135, looseness=0.9](-5.7,1)  node[bazaar] at (-1.8,3)  {$\varrho$}; 

\node[star,star points=7,star point ratio=0.8,draw, ultra thick,darkgoldenrod] at (2.6,-0.1) {\textcolor{darkgoldenrod}{$u_1$}};
\draw[operation,mint] (1.35,1.45) -- (0.9,0.45) node[mint] at (1.4,0.8)  {$\sigma$};
\node[regular polygon,regular polygon sides=9,draw, ultra thick,darkmagenta] at (-8.9,-0.65) {\textcolor{darkmagenta}{$u_2$}};
\draw[operation,paleviolet-red] (-8.45,-0.6) -- (-7,-0.6) node[paleviolet-red] at (-7.7,-0.4)  {$\lambda$};
\node[regular polygon,regular polygon sides=7,draw, ultra thick,darkcoral] at (1.6,2) {\textcolor{darkcoral}{$\text{PK}$}};
\draw[operation,copper] (2.4,0.3) -- (1.8,1.5) node[copper] at (2.3,0.9)  {$b$};
\draw[operation,midnightgr] (1.9,2.5) to[out=85,in=95, looseness=4] (1.3,2.5) node[midnightgr] at (1.6,2.85)  {$a$};
\end{tikzpicture}}
\caption{\footnotesize Schematic representation of the different interactions in model (\ref{Model2}), between tumor,  immune system and  drug dosages.}
\label{Model_Scheme}
\end{figure}
\begin{table}[!h]
\begin{center}
\caption{Numerical values and definitions of the parameters used in model~(\ref{Model2}) }\label{Table2}
\begin{tabular}{ccc}
\toprule
Parameter & Definition & Numerical value \\
\midrule
$\mu_C$ & tumor growth rate & 1.0078 $\cdot10^7$ cells/day \\
$\mu_I$ & tumor stimulated & 0.0029 day$^{-1}$ \\
& proliferation rate&\\
$\alpha$ & rate of immune  & 0.0827 day$^{-1}$\\
&cells influx&\\
$\beta$ & inverse threshold & 0.0040\\
$\gamma$ & interaction rate & 1 $\cdot10^7$ cells/day \\
$\chi$ & death rate & 0.1873 day$^{-1}$\\
$\sigma$ & chemotherapeutic  & 1  $\cdot10^7$ cells/day\\
&killing parameter&\\
$\lambda$ & immunotherapy  & 1  $\cdot10^7$ cells/day\\
&injection parameter&\\
$x_{\infty}$ & fixed carrying capacity & 780 $\cdot10^6$ cells\\
$\varrho$ & chemo-induced loss  &1\\
&on immune cells&\\
$a$ & chemotherapy  &0.5\\
&concentration decay&\\
$b$ & drug rate effect&1\\
&on the concentration&\\
&of chemotherapy&\\
\bottomrule
\end{tabular}
\end{center}
\end{table}

\begin{figure}
\centering
\includegraphics[width=0.42\textwidth]{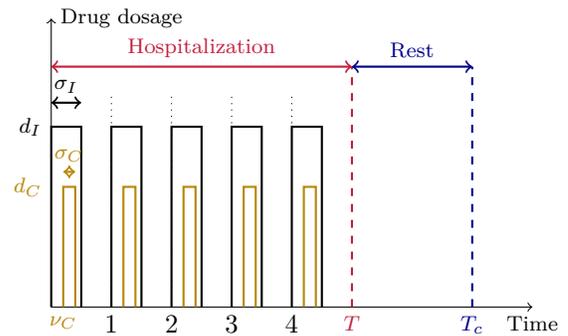}
\caption{\footnotesize Temporal open-loop control structure for each cycle, in black and yellow, respectively, the immunotherapy and the chemotherapy profiles. }
\label{Ol_control}
\end{figure}

Let's denote by $x=\left( x_1,x_2,x_3\right)$  and $u=\left( u_1,u_2 \right)$ respectively, the state and the control input vectors. In this paper, we consider a cycle-based  treatment, where the drugs are injected following   $N_{C}$ therapeutic cycles, each cycle has two phases, a hospitalization period where the drugs are injected for 5 consecutive days and a rest period where the patient recovers.  Fig.~\ref{Ol_control} shows a typical temporal 
combined control structure where the time unit is in days, $\sigma_I$  and $d_I$ 
stand for the duration and the concentration level of the immunotherapy 
injection, respectively.  The chemotherapy is assumed to be delayed from the 
immunotherapy by $\nu_C$ and is injected for $\sigma_C$ days with a 
concentration $d_C$. $T$ stands for the time of the hospitalization period 
while $T_c$ denotes the cycle duration.  Therefore, for a given treatment cycle, the therapeutic profile considered in this paper is completely defined by the following control parametrization $\theta$:

\begin{equation}
\centering
\theta=[\nu_C,\sigma_C,d_C,\sigma_I,d_I].    
\end{equation}
 
\begin{figure}[h!]
\centering
\includegraphics[width=0.45\textwidth]{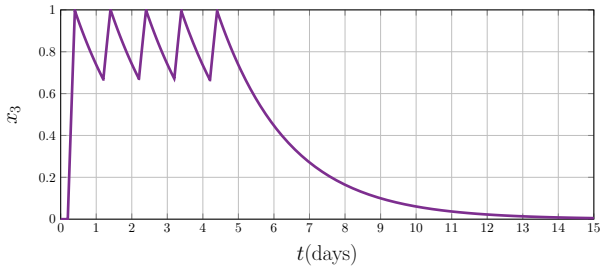}
\caption{\footnotesize A typical PK curve for chemotherapy, where $u_1$ represents 5 consecutive doses of $4.8h$, during the 5 first days of the therapy period, at a rate of one dose per day.}
\label{PK}
\end{figure}

 In cancer treatment 
design, we usually have many constraints to satisfy, they can 
be defined either on the states or on the control inputs. These constraints enable to prevent from drug toxicity and immune weakening. Therefore, we consider the following constraints:
\begin{equation}
\left\{
    \begin{array}{llll}
       x_2(t) \geq c,\hspace*{1.5mm} \forall t \in \left[0,T\right] \hspace*{1.5mm} 
\text{with}\hspace*{1.5mm} c,T \in \mathbb{R}_+,\\
        0\leq x_3 (t)\leq1,\hspace*{1.5mm} 0 \leq u_2 (t) \leq 1, \\
    \end{array}
\right.
\label{Constraints}
\end{equation}
where the first constraint is a health constraint on the minimal density of immune cells. The  constraints on $x_3(t)$ and $u_2(t)$ for all $t$,  are drug toxicity constraints. The constraint on $x_3$ can be satisfied by properly choosing a constraint on $u_1$, given the PK parameters ($a$ and $b$) since these two variables are linked through a simple first order dynamics. Fig.~\ref{PK} shows a typical PK curve, where 5 consecutive doses of chemotherapy are injected, at a rate of 1 dose per day, each dose lasts $4.8h$. We can notice that thanks to a proper choice of the constraint on $u_1$, the constraint on $x_3$ is satisfied even for successive drug doses injections. Furthermore, the constraints on the control inputs, $u_1$ and $u_2$, can be satisfied by properly choosing the parametrization $\theta$ of the control input $u$. Therefore, we will consider only the first constraint on $x_2$, since the satisfaction of the other constraints can be monitored by a proper choice of $\theta$.

The uncontrolled model (\ref{Model2}) (for $u=(0,0)$) has two locally asymptotically stable equilibria. The macroscopic malignant equilibrium is $x_m=(766.4,0.018,0)$ and the benign one is $x_b=(41.45,0.954,0)$.  The objective of  the treatment is to drive the state initial conditions to the region of attraction of the benign equilibrium (safe region), without constraints violation. Therefore,  we are interested in characterizing  the set of initial conditions (tumor volume and immune density) from which the trajectories of~({\ref{Model2}}) can be driven to the safe region under parametric uncertainties.



In this paper we aim at computing a sequence of sets 
$\displaystyle \left\{\Omega_{k}\right\}^{N_C}_{k=1} $, for  $N_C$ therapeutic cycles. Those sets are determined in   the space of the cancer burden and the ECs density, such that, in the family of  control parametrizations that we consider, there exists a therapeutic protocol  that drives, with a desired probability, the  states from $\Omega_{k+1}$ to $\displaystyle \bigcup_{j=0}^{k} \Omega_j$ without safety constraints violations. 
The set $\Omega_0$ is defined here as a probabilistically 
certified region of attraction of the benign equilibrium, when no control is applied \textit{i.e.} $u=0$. We denote by $\Omega_N$  an estimation of the region of attraction of the benign equilibrium for $u=0$, when nominal model parameters (in Table~\ref{Table2}) are considered.
Therefore, we propose a feedback strategy that can be seen in an 
implicit way, such that at the end of each therapy period, we measure the states 
(patient health and tumor volume) and depending on the certified set $\Omega_k$ where this 
measure lies, we can estimate the maximal possible recovery time ($T_c-T$) 
that the patient can take. At the end of the rest period, the certified therapy 
corresponding to this set is then applied, we keep doing this process until we 
reach the safe region $\Omega_{0}$.

\section{RoA probabilistic certification using randomized methods}
In this section, we will establish a framework of  RoA probabilistic 
certification, based on the randomized methods presented in  \cite{Alamo2009} 
and \cite{Alamo2015}. Therefore,  we will briefly recall the main key-points of 
the randomized approaches that are important for our RoA certification 
framework. The Randomized methodology  had been used to certify feedback control 
strategies in \cite{Alamir2015} for a combined cancer therapy model. In this 
paper, we propose to use this general framework in order to probabilistically 
certify the existence of a control structure which allows to drive initial 
states from a given   set to a target set under parametric uncertainties.

Let's rewrite system~(\ref{Model2}) into the following form:
\begin{equation}\label{G_sys}
 \dot{x} = f(x,u,p),
\end{equation}
where $p$ is the vector of parameters that model~(\ref{Model2}) involves. Furthermore, we consider that the variables of system~(\ref{G_sys}) are subject to the following constraints:
\begin{equation}
    x \in \mathbb{X}, \quad x(T)\in \Omega, \quad u\in \mathbb{U}.
    \label{G_constraints}
\end{equation}
In this paper, we are interested in specific control structures, since in 
cancer treatment, control inputs cannot be free real variables. They are 
usually defined by specific cyclic protocols, as illustrated in the subsequent section. 
Furthermore, in order to solve the optimization problem that we will define in the sequel, using the randomized methods, we need to consider that the control inputs
are parametrized by a vector $\theta$ which lies in a discrete set $\Theta$ 
with cardinality $n_{\Theta} \in \mathbb{N}$. However, this choice of $\theta$ remains interesting in the case of cancer therapy design, since some of the parameters involved in the treatment scheduling are naturally quantified.

We consider that the parameters vector $p$ is a random variable  following the  probability distribution
$\mathcal{P}$ that we denote $p \sim \mathcal{P}$ . Given a set $\Gamma \subseteq \R^n$ (to be more precise, $\Gamma$ must belong 
to the $\sigma$-algebra defined on $\R^n$) and a parameterization of the input 
$\theta \in \Theta$, let's consider the following optimization problem: 
\begin{equation}
\begin{aligned}
 \underset{\theta \in \Theta}{\text{min}} &\hspace{1mm} J(\theta) &  \text{s.t.} \hspace{1mm}& \forall \left(x_0, p \right) \in \left(\Gamma \times \mathbb{P}\right) & 
& g(\theta,x_0,p)=0,
\end{aligned}
\label{H_opt_prob}
\end{equation}
where $J(\theta)$ is a cost function to be minimized. In terms of cancer treatment design, this function can be a combination of many objectives that one seeks to achieve, for example reducing the quantity of injected drugs, to prevent from toxicity, or reducing the duty cycle in order to reduce the hospitalization duration. Whereas $g$  is the 
failure indicator function,  defined on the state trajectories of (\ref{G_sys}). The function $g$, in particular, is a deterministic function that, for given 
initial state, input parameter $\theta$ and model parameter $p \in \mathbb{P}$, has 
value equal to one if constraints (\ref{G_constraints}) are violated, zero otherwise. 
Problem~(\ref{H_opt_prob}), then, aims at selecting the optimal control strategy 
such that no specification violation occurs. 


The randomized method consists of replacing the original  problem 
in~(\ref{H_opt_prob}) by the following chance-constrained problem allowing some violations: 
\begin{equation}
\begin{aligned}
 \underset{\theta \in \Theta}{\text{min}}&\hspace{1mm} J(\theta)&  \text{s.t.} \hspace{1mm}   & \pr_{\mathcal{X}_0(\Gamma) \times \mathcal{P}} \left\{g(\theta,x_0,p)=1\right\} \leq \eta, 
\end{aligned}
\label{S_opt_prob}
\end{equation}
where the constraint is on the probability of constraints violation, with 
respect to the  distribution of $x_0$ on $\Gamma$, that we denote $\mathcal{X}_0(\Gamma)$, and $p \sim 
\mathcal{P}$. This problem gives therefore a chance constrained formulation in 
the sense that we can accept a vector $\theta$ which minimizes the cost $J$, 
even if the specifications are violated for some realizations of 
$\left(x_0,p\right)$, provided that the probability of these violations is lower than $ 
\eta$ hence small enough.

Since problem~(\ref{S_opt_prob}) is hard to solve, it can be simplified into the following problem, employing the empirical mean instead  
of the probability of the constraints violation: 
 \begin{equation}
\begin{aligned}
 \underset{\theta \in \Theta}{\text{min}} \hspace{1mm}& J(\theta)&  \text{s.t.} \hspace{1mm}  & \sum_{i=1}^{N} g\left(\theta,x_0^{(i)},p^{(i)}\right) \leq m,\\
& & & \left(x_0,p\right)^{(i)} \sim \left( \mathcal{X}_0(\Gamma) \times \mathcal{P}\right), \hspace{1mm} \forall i=1,\ldots,N,
\end{aligned}
\label{EM_opt_prob}
\end{equation}
where $m$ is the maximum number of constraints violation.

\begin{theorem}
For given $\Gamma \subseteq \R^n$, let $m \in \mathbb{N}$ be any integer,  let $\delta \in (0,1)$ be a targeted 
precision parameter, and suppose that problem (\ref{EM_opt_prob}) has a 
solution, that we denote $\hat{\theta}$, for $N$ i.i.d. samples of $(x_0,p)$,  with $N$ satisfying the following condition from \cite{Alamo2015}:
$$
N \geq \dfrac{1}{\eta} \left(m + \ln\left( \dfrac{n_{\Theta}}{\delta} \right) + 
\left(2m \ln\left(\dfrac{n_{\Theta}}{\delta}\right) \right)^{\frac{1}{2}} 
\right)
$$ 
Then the solution $\hat{\theta}$ satisfies the constraint in problem (\ref{S_opt_prob}) with a probability higher than $1 - \delta$.
\end{theorem}

 It is interesting to notice that the bound on $N$ in Theorem~1 provided by \cite{Alamo2015} does not depend on the dimension of the vector $(x_0, 
p)$ which is useful when having many uncertain parameters and initial states in 
the certification problem. Furthermore, the confidence parameter $\delta$ 
affects the bound with a logarithmic term which means that we can have a highly 
confident certification with a tractable number of random samples.  

Therefore, the iterative resolution of problems of the type 
(\ref{EM_opt_prob}) allows one to generate a sequence of sets $\displaystyle \left\{\Omega_{k}\right\}^{N_C}_{k=1} $ such 
that the constraints violation on passing from $\Omega_{k+1}$ to $\displaystyle \bigcup_{j=0}^{k} \Omega_j$ 
is smaller then $\eta$ with a certain desired confidence probability $1-\delta$.  

\subsection{Algorithm for RoA estimation}
Given a target set $\Omega$, let's suppose that our objective is to certify that the set $\Gamma$ is such that there exists a control parametrization $\theta$, for which at least $100\cdot (1-\eta)\%$ of the trajectories of  (\ref{G_sys}), generated by the distributions of the initial states $x_0 \in \Gamma$ and the uncertain parameters $p$,  converge to $\Omega$ at time $T$, while satisfying constraints (\ref{G_constraints}), with a probability of confidence higher than $1-\delta$. 
Any solution of (\ref{EM_opt_prob}) defines a local control strategy that satisfies the constraints while minimizing the cost $J(\theta)$. 


\paragraph*{\bf{$\Gamma$ generator}}
we suppose that we have a generator of sets $\Gamma$ with a parametrized geometry  providing a family of nested potential sets $\Gamma$, then we can compute the biggest one that is probabilistically certified through ({\ref{EM_opt_prob}}). 

Therefore, starting from $\Omega_0$ which is known to be in the region of attraction of the desired equilibrium, an iterative procedure can be designed to generate the sequence $\displaystyle \left\{\Omega_k\right\}_{k=0}^{N_C}$ such that the trajectories starting in $\Omega_{k+1}$ end in $\displaystyle \bigcup_{j=0}^{k} \Omega_j$ with the desired probability and  without violating the constraints. In particular, we will consider sequences of sets such that $\Omega_{k} \cap \Omega_{k+1} = \emptyset$.
Then we keep doing this certification process until, given $\Omega_{k-1}$, the set $\Omega_{k}$ is empty. Once the RoA probabilistic certification algorithm terminates, the probabilisitically certified RoA is the set $\Omega_{C} = \bigcup\limits_{i=1}^{N_C} \Omega_i$. 

Note that, if $x_0 \in \Omega_{k}$ for $k = 1, \cdots, N_C$, this means that the trajectory of length $T$ will end in $\displaystyle \bigcup_{j=0}^{k-1} \Omega_j$ without violating the constraint with a certain probability, but no direct probabilistic guarantee is given regarding the convergence to the set $\Omega_0$. It is not straightforward to derive a probabilistic bound on driving the states directly from the last set of the sequence $\Omega_{N_C}$ to $\Omega_0$. This because it involves the accuracy and confidence parameters, $\eta$ and $\delta$, but also since there is no guarantee that, given the initial state distribution $\mathcal{X}_0(\Omega_k)$, the distribution of the state at the end of the $k$-th therapeutic cycle 
is $\mathcal{X}_0(\Omega_{k-1})$, for which the probabilistic validation is performed. However, after deriving the sequence of certified sets, we can approximate the probability of driving the states from $\Omega_{N_C}$ to $\Omega_0$, with  the derived certified control strategy, using Monte-Carlo simulations.


\begin{algorithm}[!h]
\caption{Sequence of probabilistically certified sets}
\algorithmicrequire $\hspace{2mm} \Omega_0$
\begin{algorithmic} 
\STATE $k \leftarrow 0$	
\WHILE{ $\Omega_k \neq \emptyset$}
\STATE{$ \displaystyle \Omega \leftarrow \bigcup_{j=0}^{k} \Omega_j$} 
\REPEAT 
\STATE{\text{Generate} $\Gamma$} 
\UNTIL{(\ref{EM_opt_prob}) is unfeasible for $\Gamma$}
\STATE $k \leftarrow k+1$	
\STATE $\Omega_k\leftarrow \Gamma$
\ENDWHILE
\STATE $N_C \leftarrow k-1$\\
\end{algorithmic}
\algorithmicensure $\hspace{2mm}  \displaystyle \Omega_{C} \leftarrow \bigcup_{i=0}^{N_C} \Omega_i$
\label{Algo_cert}
\end{algorithm}
Finally, by using Algorithm~\ref{Algo_cert}, we can obtain a sequence of certified sets, such that the output is the probabilistically certified RoA $\Omega_C$. 

\section{Probabilistically certified RoA for cancer model}

In cancer treatment, therapies are usually injected following many successive cycles with specific periods of hospitalization and rest. This makes the certification framework presented in the previous section suitable to certify RoAs for a cancer model. Hence, considering $N_C$ treatment cycles, our objective consists of estimating the probabilistically certified RoA of model~(\ref{Model2}) that we denote $\Omega_C$. To this end, we certify a sequence of successive disjoint sets such that their union is the probabilistically certified RoA. Moreover, the temporal control profiles that we consider correspond only to the hospitalization period (see Fig.~\ref{Ol_control}), meaning that the rest period is not included in the decision variable $\theta$ defined in Section~II, since we assume that this parameter can be estimated afterwards depending on the health conditions of the patient.

The initial condition $x_0$ is assumed to be uniformly distributed in the set $\Gamma$ while the parameters of model~(\ref{Model2}) are assumed to be normally distributed in the intervals $\left[0.9p_{nom},1.1p_{nom}\right]$, where $p_{nom}$ is the nominal value of each parameter and the variance of these distributions is 0.01. The parameter $x_{\infty}$ is supposed to be known and doesn't follow any distribution.

The failure indicator function, which indicates whether the constraints 
(\ref{Constraints}) are satisfied or not, is defined on $x(t|x_0,p,\theta)$  which is the state trajectory of (\ref{Model2}) for a given control parametrization $\theta$ and a random sample of $x_0$ and $p$. We denote by $x(T|x_0,p,\theta)$ the state trajectory $x$ evaluated at the end of the hospitalization period. Therefore, the failure indicator is defined as:
$$
g(\theta,x_0,p,\Omega) := \left\{
    \begin{array}{lll}
        0 & \: \: \: \mbox{if } x_2(t|x_0,p,\theta) \geq c \hspace*{2mm} 
\forall 
t \hspace*{0mm} \\
        & \hspace*{3mm} \text{and} \hspace*{2mm}x(T|x_0,p,\theta) \in \Omega \\
        1 & \: \: \: \mbox{otherwise}
    \end{array}
\right. 
$$
where $\Omega$ is a  probabilistically certified target set which can be seen 
as 
the safe region.  
Using Algorithm~\ref{Algo_cert}, we can derive a sequence of probabilistically certified sets providing the probabilistically certified RoA.  Firstly, we need to derive an initial target set $\Omega_0$, in order to initialize the certification algorithm. 

\subsection{Probabilistically certified initial target set $\Omega_0$}

We define the probabilistically certified initial target set $\Omega_0$ as being the  uncontrolled probabilistically certified region of attraction of many locally asymptotically stable equilibriums. Therefore, given  $p \in \mathbb{P}$ (drawn according to the probability distribution $\mathcal{P}$) and $x_0$ following  a uniform distribution on $\Omega_0$, that we denote $\mathcal{U}(\Omega_0)$, we certify that:
\begin{equation}
\footnotesize
\pr_{\mathcal{U}(\Omega_0) \times \mathcal{P}} \left\{x_2(t|x_0,p) \geq c, \hspace{1mm} \forall t>0 \hspace*{1mm} \wedge \hspace*{1mm}  x(T|x_0,p) \in \Omega_{eq} \right\} > 1- \eta,
\label{Omega0}
\end{equation}
for a given time $T$. We denote by $\Omega_{eq}$ a certified set in a neighborhood of benign equilibriums of (\ref{Model2}), generated by the realizations of $p$ according to the probability distribution $\mathcal{P}$. Therefore $\Omega_{eq}$  is derived such that:
\begin{equation}
\footnotesize
\pr_{\mathcal{U}(\Omega_{eq}) \times \mathcal{P}} \left\{x_2(t|x_0,p) \geq c, \hspace{1mm} \forall t>0 \hspace*{1mm} \wedge \hspace*{1mm}  x(T|x_0,p) \in \Omega_{eq} \right\} > 1- \eta,
\label{Omega_eq}
\end{equation}
Note that $\Omega_{eq}$ is slightly different than a probabilistically certified invariant set, since we don't require that the trajectories starting in $\Omega_{eq}$ stay in it, we rather require that these trajectories satisfy the constraints~(\ref{Constraints}) and converge to $\Omega_{eq}$ after some time $T$. Moreover, the set $\Omega_{eq}$ is derived in order to be used as a target set for the determination of $\Omega_0$.

\begin{figure}[h]
 \centering
\includegraphics[width=0.38\textwidth]{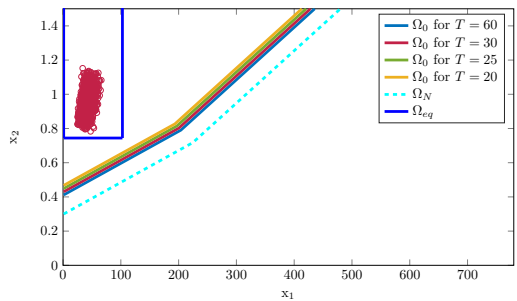}
	\caption{\footnotesize Probabilistically certified sets $\Omega_0$ for different times T}
  	\label{RoAs_Cert_Without_u}
\end{figure}

In order to find $\Omega_{eq}$, we  draw the distribution of the benign equilibriums of model~(\ref{Model2}) when it is subjected to parametric uncertainties.  Then, we choose a geometry for $\Omega_{eq}$ surrounding the benign equilibriums of the sample shown in Fig.~\ref{RoAs_Cert_Without_u}. Finally, we expand this set until (\ref{Omega_eq}) is not satisfied.
Fig.~\ref{RoAs_Cert_Without_u} shows the probabilistically certified RoA of benign equilibriums $\Omega_{eq}$, the estimated uncontrolled nominal region of attraction $\Omega_N$ and the initial probabilistically certified target set $\Omega_0$ for different $T$. 

Using the phase-portrait of system~(\ref{Model2}) with $u=0$ and considering nominal parameters (in Table~\ref{Table2}), we give an estimate of the nominal region of attraction of the benign equilibrium for system~(\ref{Model2}) without control, denoted $\Omega_N$, see Fig.~\ref{RoAs_Cert_Without_u}. After finding a proper geometry for the set $\Omega_{eq} $ such that it satisfies (\ref{Omega_eq}), we use the transition between (\ref{S_opt_prob}) and (\ref{EM_opt_prob}) provided by the randomized methods, in order to certify the set $\Omega_0$. Note that in this case $\mathcal{X}_0(\Gamma)$ corresponds to $\mathcal{U}(\Omega_0)$ since we assume that $x_0$ is uniformly distributed on $\Omega_0$, and the target set for the states at time $T$ denoted  $\Omega$ in the definition of $g$ corresponds to $\Omega_{eq}$. Furthermore, since we deal with an uncontrolled problem, we have that $\theta=0$, therefore, (\ref{S_opt_prob}) turns out to be a feasibility problem, where we need only to guarantee the probability condition  in (\ref{Omega0})  by using the empirical mean over $g$ for $N$ i.i.d. samples of $(x_0,p)$ mentioned in (\ref{EM_opt_prob}), with $\theta=0$ and $n_{\theta}=1$, the bound $N$ is then given by Theorem~1.


Therefore, we assume that the set $\Omega_0$ to be certified has the same geometry as the estimated nominal uncontrolled region of attraction $\Omega_N$ that we shrink  until (\ref{Omega0}) is not satisfied given the confidence probability $1-\delta$. There is clearly no guarantee that the set $\Omega_0$ that we obtain is the biggest possible certified set, however, in this case, proving the existence of a set $\Omega_0$ satisfying (\ref{Omega0}) is enough, since $\Omega_0$ is only used as a target set for the Algorithm~\ref{Algo_cert} allowing therefore to compute the sequence of certified sets.

\subsection{Probabilistically certified  region of attraction $\Omega_C$}
We denote by $\Omega_C$ the  probabilistically certified region of attraction of system (\ref{Model2}). We initialize Algorithm~\ref{Algo_cert} with $\Omega_0$ in order to derive the sequence of probabilistically certified sets providing the certified RoA for model~(\ref{Model2}).
We consider that the decision variable $\theta$ is defined by the following variables:
\begin{equation*}
\left\{
    \begin{array}{llll}
\sigma_I \in \{0,0.16,0.32,0.48,0.64, 0.8\},\\
\sigma_C=0.2,\quad \nu_C =0.2,\\
d_I \in \{ 0,0.25, 0.5,0.75,1\},\\
\bar{d}_C \in \{0,0.11,0.22,0.33,0.44, 0.56,0.67,0.78,0.89,1\},\\
    \end{array}
\right.
\label{Consider}
\end{equation*}
where $\bar{d}_C$ denotes the maximal desired concentration of $x_3$ allowing to monitor $d_C$. Therefore, the cardinality of $\Theta$ is  $n_{\Theta}=300$ giving the bound $N \geq 1863$ according to Theorem~1, for $\eta=10^{-2}$ and $\delta=10^{-3}$. The number of simulations to be performed for each set certification is $N_{sim}=N \cdot n_{\Theta}= 558900$. The  required computational time to perform $N_{sim}$ simulations is around $5.79mn$ using Matlab coder toolbox, therefore, 1 simulation requires around $621 \mu s$ on an hp EliteBook 2.60GHz Intel Core i7.

\begin{figure}[h]
	\centering
\includegraphics[width=0.38\textwidth]{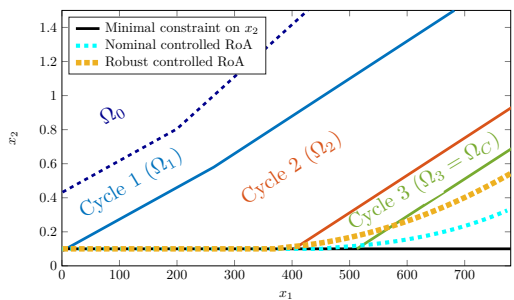}
	\caption{\footnotesize Probabilistically certified RoAs for 3 injection cycles}
  	\label{DA_p22}
\end{figure}

Fig.~\ref{DA_p22} shows the 3 certified cycles for $T=5$ obtained using Algorithm~\ref{Algo_cert}, nominal and robust RoAs that have been estimated using a sliding-mode-based method, where bang-bang feedback control is considered. We can see that, as the number of cycles increases, the certified RoA gets closer to the robust controlled one. Furthermore, it is interesting to notice that there is a small region of $\Omega_3$, which is probabilistically certified but does not belong to the robust RoA, although the control structure in the robust case is less restrictive. This is potentially due to the fact that the probabilistic method is less conservative than the robust one.
Furthermore, we approximated the probability of driving the states from $\Omega_3$ to $\Omega_0$ using 5000 Monte-Carlo simulations. We obtained that $99.6\%$ of the trajectories of ({\ref{Model2}}) having initial conditions in $\Omega_3$ converge to $\Omega_0$ using the probabilistic certified control strategies that we derived.
\section{Conclusion}
In this paper, we presented a framework of probabilistic certification for regions of attraction. This framework is based on the randomized methods which allow to  overcome the conservatism of worst-case robust approaches by proposing a tractable problem with probabilistic constraints.
The framework of region of attraction certification can be seen as a tool to tune the several parameters of treatment protocols by properly choosing the model parameters and their distributions, the geometry of the regions of attraction to be certified and the control parametrization.   
An interesting perspective for future work would be to apply this methodology to other systems describing cancer dynamics. 
\bibliography{Biblio}
\bibliographystyle{abbrv}

\end{document}